\documentstyle[12pt,epsfig]{article}

\def\beq{\begin{equation}}
\def\eeq#1{\label{#1}\end{equation}}
\def\eeqn{\end{equation}}
\def\beqa{\begin{eqnarray}}
\def\eeqa#1{\label{#1}\end{eqnarray}}
\def\eeqan{\end{eqnarray}}

\def\leqn#1{(\ref{#1})}

\def\to{\rightarrow}

\def\stacksymbols #1#2#3#4{\def\theguybelow{#2}
    \def\vp{\lower#3pt}
    \def\sp{\baselineskip0pt\lineskip#4pt}
    \mathrel{\mathpalette\intermediary#1}}

\def\intermediary#1#2{\vp\vbox{\sp
     \everycr={}\tabskip0pt
     \halign{$\mathsurround0pt#1\hfil##\hfil$\crcr#2\crcr
              \theguybelow\crcr}}}

\begin{document}

\begin{titlepage}

\vskip 5cm
\begin{center}
{\Huge \bf Dark Matter Identification with}
\vskip0.5cm
{\Huge \bf Gamma Rays from Dwarf Galaxies} 

\vskip.2cm
\end{center}
\vskip1cm

\begin{center}
{\bf Maxim Perelstein and Bibhushan Shakya} \\
\end{center}
\vskip 8pt

\begin{center}
	{\it Institute for High Energy Phenomenology\\ 
	Newman Laboratory of Elementary Particle Physics\\
	     Cornell University, Ithaca, NY 14853, USA } \\

\vspace*{0.3cm}

{\tt  mp325@cornell.edu, bs475@cornell.edu}
\end{center}

\vglue 0.3truecm

\begin{abstract}
If the positron fraction and combined electron-positron flux excesses recently observed by PAMELA, Fermi and HESS are due to dark matter annihilation into lepton-rich final states, 
the accompanying final state radiation (FSR) photons may be detected by ground-based atmospheric Cherenkov telescopes (ACTs). Satellite dwarf galaxies in the vicinity of the Milky Way are particularly promising targets for this search. We find that current and near-future ACTs have an excellent potential for discovering the FSR photons from dwarfs, although a discovery cannot be guaranteed due to large uncertainties in the fluxes resulting from lack of precise knowledge of dark matter distribution within the dwarfs. We also investigate the possibility of discriminating between different dark matter models based on the measured FSR photon spectrum. For typical parameters, we find that the ACTs can reliably distinguish models predicting dark matter annihilation into two-lepton final states from those favoring four-lepton final states (as in, for example, ``axion portal" models). In addition, we find that the dark matter particle mass can
also be determined from the FSR spectrum.  
\vskip 3pt \noindent

\end{abstract}

\end{titlepage}

\section{Introduction}

Recent measurements of the positron fraction in cosmic rays in the 10--80 GeV range by PAMELA~\cite{pamela} and the combined electron-positron flux up to the TeV scale by Fermi~\cite{fermi} and HESS ~\cite{hess} indicate excesses inconsistent with conventional astrophysical background from cosmic ray nuclei interactions with the interstellar medium. Although the possibility that the origin of these excesses could lie within conventional astrophysics, such as nearby pulsars~\cite{pulsars}, remains, there is widespread optimism that these excesses could be the first indirect evidence of dark matter annihilation in the galaxy. 

The simple hypothesis that dark matter consists of weakly interacting massive particles (WIMPs) yields the correct current dark matter density, and is supported by many popular extensions of the standard model of particle physics. It can also fairly naturally explain these excesses. The WIMP framework requires the dark matter particles to have a (thermally averaged) annihilation cross-section of $\langle \sigma v \rangle\sim3\times10^{-26}$ cm$^3$s$^{-1}$ at freeze-out, and a mass in the range from 10 GeV to 10 TeV. PAMELA, Fermi, and HESS excesses can be explained, in a way consistent with all other astrophysical observations, by dark matter with an annihilation cross section of $\langle \sigma v\rangle \sim3\times10^{-23}$ cm$^3$s$^{-1}$ in the Milky Way, and a mass of 1--3 TeV, provided annihilation is predominantly into electrons or muons~\cite{patrick,bestfits}.\footnote{The most significant constraint on these models currently arises from the first-year Fermi-LAT diffuse gamma ray map~\cite{fermi_ics}. 
An isothermal-like dark matter density profile in the Milky Way is required to satisfy this constraint.
The constraint is especially important for the muon final states~\cite{CC}.}
The $\mathcal{O}$(100-1000) discrepancy in annihilation cross section can be bridged by incorporating Sommerfeld enhancement~\cite{sommerfeld,dmtheory} and/or a ``boost factor" associated with small-scale dark matter clumps in the Milky Way halo\footnote{The degree of clumps that would provide all of the required enhancement, however, is disfavored by numerical simulations~\cite{clumps}.}~\cite{clumps}. 

If the observed excesses indeed have a dark matter origin, accompanying signals are expected in the form of energetic gamma rays. For leptophilic dark matter, these are produced in two ways:  inverse Compton scattering (ICS) of starlight and CMB photons off energetic leptons from dark matter annihilation, and final state radiation (FSR) from the annihilation process~\cite{700,dmseeslight,lightparticleconstraint,muonspectrum,patrick,rouven}. We will focus on the FSR photons in this paper. The FSR component is dominant at energies close to the dark matter mass. It is also independent of most of the astrophysical uncertainties that plague the calculation of positron and ICS photon fluxes. Finally, if WIMP annihilation primarily results in two-body leptonic final states, the associated FSR spectrum has a sharp ``edge" feature at the energy equal to the WIMP mass, enhancing its visibility over astrophysical backgrounds and allowing for precise mass measurement if an edge is observed~\cite{robust}. 

Dark matter annihilation rate, and with it the FSR photon flux, are proportional to the square of the dark matter density, and regions with enhanced density should provide the strongest flux. The largest concentration of dark matter in our cosmic neighborhood is of course in the central region of the Milky Way. 
However, fluxes of hard gamma rays from astrophysical sources located near the galactic center are large and poorly understood, providing a serious background to dark matter searches in that region. In contrast, dwarf galaxy satellites of the Milky Way, which are highly dark matter dominated, are largely free of astrophysical backgrounds, and many of them lie away from the galactic center~\cite{dwarfsgeneral}. This led many authors to advocate dwarfs as promising sources for gamma ray signals of dark 
matter~\cite{dwarfdm,magic2ctaprospects,rouven}. 

The Fermi Large Area Telescope (LAT) can cover the whole sky continuously, detect photons down to 100 MeV energy, has superior energy and angular resolution, and is free of atmospheric background. However, as dark matter gamma ray signals from dwarf galaxies are mainly constrained by low statistics, atmospheric Cherenkov telescopes (ACTs) offer a distinct advantage:
the typical ACT effective area is $\mathcal{O}$(10$^9$) cm$^2$, versus the $\sim 10^4$ cm$^2$ effective area of Fermi, offsetting the disadvantages of limited observation time, poorer resolution, and atmospheric background.\footnote{Fermi LAT may be used to constrain properties of dark matter with gamma rays originating from the galactic center, see Ref.~\cite{FermiMD}.} An additional shortcoming of the ACTs in many searches is their high energy threshold, typically around 200 GeV. However, as the FSR photon spectrum extends all the way to the energy corresponding to the WIMP mass, which is required to be in the 1--3 TeV range to explain PAMELA, Fermi and HESS electron/positron anomalies, this limitation does not play a crucial role in our case. 

Observation of dwarf galaxies with ACTs for possible signals of dark matter annihilation has been carried out on numerous occasions, without any positive results so far. These searches already put some bounds on leptophilic models of dark matter, as we discuss in Section~\ref{dwarfsacts}. (Some of these bounds have also been considered in Refs.~\cite{lightparticleconstraint,rouven,sommboundact}.) However, the bounds depend sensitively on the distribution of dark matter inside the dwarfs, which is at present poorly known. As a result, none of the models we examine is currently ruled out. Future searches can benefit from recent discoveries of new dwarf galaxies by the Sloan Digital Sky Survey~\cite{sloan} -- in particular, the dwarf galaxy Segue 1~\cite{segue1} -- as well as rapid advances in the field of ACTs -- MAGIC II recently became operational~\cite{magic2}, while the Cherenkov Telsecope Array (CTA)~\cite{cta} and the Advanced Gamma-ray Imaging System (AGIS)~\cite{agis,agis2}, which will provide an order of magnitude improvement over current ACTs, are due to be completed in the next few years. We discuss the prospects of detection in Section~\ref{prospects}. We find that the ongoing and near-future ACTs will cover most of the flux range predicted by the leptophilic models, although unfortunately the large uncertainties on dark matter distribution in the dwarfs do not allow us to conclude that the discovery is guaranteed. 

Inspired by the discovery prospects, in Section~\ref{moddisc} we investigate the important question of whether the ACTs can measure the FSR photon spectrum precisely enough to distinguish between different dark matter annihilation channels predicted by different models of dark matter. We find that, for typical dark matter and astrophysical parameters, model discrimination prospects are good: in particular, the ACTs can reliably distinguish models which predict two-body final states from those favoring four-body final states. The latter possibility is common in models where annihilation proceeds via a scalar ``portal" particle; the same particle may give rise to the Sommerfeld enhancement of the cross section~\cite{dmtheory,portal}. Discrimination between the different four-body final states (4$e$ vs. 4$\mu$) is more challenging, since the FSR spectra in these two cases are quite similar. In addition, we find that spectral fits provide accurate determination of the dark matter mass and annihilation cross section. 

\section{The  Final State Radiation Spectrum}

Direct annihilation of WIMPs into photon pairs cannot occur at tree level, and thus has strongly suppressed cross section. In contrast, final state radiation (FSR) of photons can occur whenever WIMPs annihilate into electrically charged particles, {\it i.e.} in the vast majority of possible SM final states (the only exception being all-neutrino channels). Most of the energetic FSR photons are approximately collinear with the emitting charged particle. In this kinematic regime, factorization theorems ensure that the energy spectrum of the FSR photons is to leading order independent of the details of the WIMP annihilation process, so that quasi-model-independent predictions can be made.\footnote{This is analogous to the quasi-model-independent predictions for WIMP production at colliders in association with a photon or a gluon~\cite{WIMPcoll}.}  Moreover, for WIMP annihilations into two-body fermionic final states, {\it e.g.} $\chi\chi\to e^+e^-$, the FSR spectrum has a characteristic ``edge'' feature at the energy equal to the WIMP mass, providing a powerful signature to discriminate it against the poorly known astrophysical background. All these features make FSR photons a promising candidate for dark matter searches, as emphasized in Ref.~\cite{robust}.

Specifying the final state of the WIMP annihilation process completely fixes the FSR spectrum. In this paper, we will focus on the final states containing charged leptons. We will assume that there is a single WIMP particle, $\chi$, with the mass in the 1 -- 3 TeV range, that can annihilate into leptons either directly, or via an intermediate ``portal" particle $\phi$, with mass of order 1 GeV. Specifically, we consider the following three annihilation channels: 
\begin{enumerate}
\item Model A: $\chi\chi\rightarrow\mu^+\mu^-$. 
\item Model B: $\chi\chi\rightarrow\phi\phi\rightarrow4e$. 
\item Model C: $\chi\chi\rightarrow\phi\phi\rightarrow4\mu$.
\end{enumerate}
These parameters and final states are favored by the combination of PAMELA, Fermi and HESS data
with bounds on gamma ray flux from the galactic center and other astrophysical constraints~\cite{patrick,bestfits}.

For dark matter annihilating into a lepton-antilepton pair $l\bar{l}$ as in model A, the FSR flux (within the leading-log approximation)  is~\cite{robust}
\beq
\frac{d\Phi_{FSR}}{dx}=\Phi_0\left(\frac{\left<\sigma v\right>}{1pb}\right)\left(\frac{100~{\rm GeV}}{m_{\chi}}\right)^3F(x)\,\log\left(\frac{4m_\chi^2(1-x)}{m_l^2}\right)\,J,
\eeq{2state}
where $x=2E_{\gamma}/\sqrt{s}=E_{\gamma}/m_{\chi}$, $\Phi_0=1.4\times 10^{-14}$ cm$^{-2}$s$^{-1}$GeV$^{-1}$, $F(x)$ is the splitting function for fermions
\beq
F(x)=\frac{1+(1-x)^2}{x},
\eeqn
and $J$ is the dimensionless astrophysical factor\footnote{Eq.~\leqn{2state} and similar formulas below are equally applicable for decaying dark matter, with an appropriate redefinition of the $J$ factor. 
However, we do not consider decaying dark matter in this paper since the resulting FSR signals from dwarf galaxies are generally too weak to be detected. See Ref.~\cite{decay} for a discussion of discriminating between decaying and annihilating dark matter based on gamma ray observations.}
\beq
J=\frac{1}{8.5~{\rm kpc}}\left(\frac{1}{0.3~{\rm GeV/cm}^3}\right)^2L,~~~~L= \int d\Omega\int_{l.o.s.}\rho^2 dl.
\label{astrofactor}
\eeqn
For $\chi\chi\rightarrow\phi\phi\rightarrow 4l$, as in models B and C, 
the FSR spectrum is~\cite{robust}
\beq
\frac{d\Phi_{FSR}}{dx}=\Phi_0\left(\frac{\left<\sigma v\right>}{1pb}\right)\left(\frac{100~{\rm GeV}}{m_{\chi}}\right)^32\frac{2-x+2x\log x-x^2}{x}\,\log\left(\frac{m_\phi^2}{m_l^2}\right)\,J.
\eeq{4state}
For the remainder of this paper, we use $\left<\sigma v\right>=3\times10^{-23}$ cm$^3$s$^{-1}$ (corresponding to, for example, an unboosted $\sigma_0=1$ pb and a factor of 1000 from Sommerfeld enhancement), $m_{\chi}=3$ TeV, and $m_{\phi}=1$ GeV\footnote{A 1 GeV portal particle is only consistent with observations if it does not have significant couplings to quarks, since otherwise $\phi$ decaying to neutral pions produce excessive gamma-ray flux from the galactic center. We assume such a scenario in this paper. An alternative is to consider a portal particle that does couple to quarks (such as, for example, an extra U(1) gauge boson mixed with the photon), but is too light to decay to pions. In this case, decays to muons are only allowed if ${m_\phi}$ is in a narrow range about 220-250 GeV, which seems unnatural. Let us note, however, that the formalism of this section would apply to this case as well, with a possible exception of model C where $(m_{\mu}/m_{\phi})^2$ suppressed corrections would need to be included in the FSR photon spectrum.} unless stated otherwise. These parameters provide consistent fits to PAMELA, Fermi and HESS data for all three models at hand~\cite{patrick,bestfits}. While lower velocities in the dwarf galaxies can lead to greater Sommerfeld enhancement, additional constraints from CMB observations~\cite{cmb} suggest that they cannot be much larger than the value we have chosen at this mass scale. Note that the use of leading-log approximation assumed in Eqs.~\leqn{2state,4state} is well justified for these parameters: the largest correction beyond the leading-log occurs in model C, and is of order $\mathcal{O}$$((\frac{2m_{\mu}}{m_{\phi}})^2)\sim 0.04.$

For models A and C, annihilation is to muons, and final state radiation form the subsequent decay of the muon also needs to be taken into account. The FSR photon spectrum from radiative muon decay in the muon rest frame is~\cite{mugamma,muonspectrum}: 
\beq
\frac{d\Phi_{\gamma}}{dw}=\frac{\alpha}{3\pi}\frac{1-w}{w}\left((3-2w+4w^2-2w^3)\ln\frac{1-w}{r}-\frac{17}{2}+\frac{23w}{6}-\frac{101w^2}{12}+\frac{55w^3}{12}\right),
\label{muonrf}
\eeqn
where $r=m_e^2/m_{\mu}^2,~w=2E_{\gamma}/m_{\mu}$. (An approximation $r\ll 1$ has been made in this formula.) If the muons come directly from dark matter annihilation, this spectrum needs to be boosted to the dark matter rest frame:
\beq
\frac{d\Phi_{\gamma}}{dy}=2\int_y^1dw\frac{1}{w}\frac{dN_{\gamma}}{dw},
\eeqn
where $y=E_{\gamma}/m_{\chi}$.
When annihilation is through an intermediate $\phi$ particle, Eq.~\leqn{muonrf} needs to be boosted twice to get to the dark matter rest frame~\cite{rouven}: first to $\phi$ rest frame,
\beq
\frac{d\Phi_{\gamma}}{dx}=2\int_{2x/(1+\beta)}^{min(1,2x/(1-\beta))}dw\frac{1}{w}\frac{dN_{\gamma}}{dw},
\eeqn
where $x=E_{\gamma}/m_{\phi}$, followed by a boost to the $\chi$ rest frame,
\beq
\frac{d\Phi_{\gamma}}{dy}=2\int_y^1dx\frac{1}{x}\frac{dN_{\gamma}}{dx},
\eeqn
where $y=E_{\gamma}/m_{\chi}$ as before. The overall FSR signal is obtained by adding this contribution to FSR off the muons. Typically, FSR off muons from the annihilation process remains dominant unless $m_{\phi}\sim m_{\mu} $\cite{rouven}. 

\section{Dwarf Galaxies and ACTs}
\label{dwarfsacts}

In this section, we will summarize the relevant properties of the known dwarf galaxies that are potential sources of observable FSR photons, as well as the basic parameters of existing and near-future atmospheric Cherenkov telescopes (ACTs). We will also discuss the bounds placed on the models at hand by the previous ACT searches for anomalous gamma rays from dwarfs.

\subsection{Dwarf Galaxies}

Dwarf satellite galaxies, made up almost entirely of dark matter, provide excellent prospects for detecting dark matter signals. After the galactic center -- where astrophysical backgrounds are large and poorly understood -- these dwarfs have the largest dark matter density in our galactic neighborhood. In addition, they have no detected neutral or ionized gas, minimal dust, no magnetic fields, and little or no recent star formation activity \cite{dwarfsgeneral}: in other words, one should expect essentially no astrophysical background associated with the dwarfs. 

Table~\ref{dwarftable} lists the major known dwarf galaxies that are promising candidates for dark matter detection, with their astrophysical factors as defined in Eq.~\leqn{astrofactor} and their galactic coordinates. The astrophysical factors are taken from Ref.~\cite{rouven}, where they are calculated for $\Delta\Omega\approx2.2\times10^{-5}$sr, which corresponds to the angular resolution of the HESS telescope, $\Delta\theta=0.15^{\circ}$. These dwarfs have already been observed by one or more ACTs, except Segue 1, which is a recently discovered candidate~\cite{segue1} that seems to be particularly promising because of its large dark to luminous matter ratio, relative proximity, and high altitude. Segue 1 has been observed by Fermi~\cite{fermisegue} with null results. We will not use Sagittarius in our analysis as it is being tidally disrupted by the Milky Way, making the estimates of the $L$ factor unreliable. In addition, it also suffers from a larger background since its direction is close to that of the galactic center. 

\begin{table}[h!]
\begin{center}
\begin{tabular}{|c|c|c|} \hline
Dwarf & $\log_{10}(L\times$\,GeV$^{-2}$cm$^{5}$)~\cite{rouven} & galactic coordinates $(l,b)$~\cite{coordinates}\\
\hline
Sagittarius & $19.35\pm1.66$&$5.6^\circ,~-14.2^\circ$\ \\
Draco &$18.63\pm0.60$&$86.4^\circ,~34.7^\circ$\\
Ursa Minor &$18.79\pm1.26$&$105.0^\circ,~44.8^\circ$\\\
Willman 1 & $19.55\pm0.98$&$158.6^\circ,~56.8^\circ$\\
Segue 1 & $20.17\pm1.44$ & $220.5^\circ,~50.4^\circ$\\
\hline
\end{tabular}
\caption{Dwarf galaxy candidates for dark matter indirect detection.}
\label{dwarftable}
\end{center}
\end{table}
The Sloan Digital Sky Survey~\cite{sloan} has recently discovered many new dwarf galaxies. Since only a small region of the galactic neighborhood has been completely surveyed, it seems likely that several hundred more low-luminosity, dark matter dominated dwarf galaxies might still be discovered in the future~\cite{dwarfs2010}. It would be straightforward to apply the analysis of this paper to any promising new dwarf  that may be discovered, once the distribution of dark matter is mapped out to allow for at least an approximate determination of its $L$ factor.

\subsection{Atmospheric Cherenkov Telescopes}

\begin{table}[h!]
\begin{center}
\begin{tabular}{|c|c|c|c|} \hline
ACT & Effective area (cm$^2$)& energy resolution $\epsilon$ \\
\hline
MAGIC \cite{magiccrab}& $10^9$ & 0.20 \\
HESS \cite{hesstelescope}&$7\times10^8$& 0.15\\
VERITAS \cite{veritas}\cite{veritas2} & $7\times10^8$ & 0.15 \\
CANGAROO III \cite{cangaroo}&$10^9$&0.15-0.30\\
STACEE \cite{staceedraco}& $10^8$& 0.25 \\
MAGIC-II \cite{magic2}\cite{magic2simulations}& $10^9$ & 0.10-0.15\\
\hline
AGIS \cite{agis}\cite{agis2}& $10^{10}$ & ? \\
CTA \cite{cta}& $10^{10}$ & 0.10 \\
\hline
\end{tabular}
\caption{Some current and near-future ACT's relevant for dwarf galaxy observations. MAGIC II started taking data in 2009. CTA is expected to be operational by 2015. Construction of AGIS could start between 2011-2013.}
\label{telescopes}
\end{center}
\end{table}

Since gamma ray fluxes from FSR in dark matter annihilation in dwarf galaxies are rather weak, the key parameter governing the sensitivity of a given telescope in this search is its effective area $A_{\rm eff}$.
In addition, an important ACT parameter is the energy resolution of the instrument $\epsilon$. The probability to assign an energy $E$ to a primary gamma ray of true energy $E^\prime$ can be approximated with a Gaussian as~\cite{magic2ctaprospects}
\beq
R_{\epsilon}(E-E')\approx \frac{1}{\sqrt{2\pi} E' \epsilon}\,\exp\left(-\frac{(E-E')^2}{2\epsilon^2E'^2}\right)\,.
\label{Eres}
\eeqn
This parameter is particularly important for measuring the gamma ray spectrum, which can be used to identify the dark matter model if an FSR signal is observed (see Section~\ref{moddisc}).
Table~\ref{telescopes} lists current and future ACT's relevant for indirect detection of dark matter. The listed effective area values are averages in the several hundred GeV to 1 TeV range; more detailed plots of effective area as a function of energy for individual instruments can be obtained from the respective references. Instead of using telescope-specific, energy dependent effective areas, we ignore energy dependence and use typical parameters in this paper so that our estimates remain more generally applicable. Unless stated otherwise, we assume an angular resolution of $\Delta\theta=0.15^{\circ}$, corresponding to a solid angle coverage $\Delta\Omega\approx2.2\times10^{-5}$ sr. The typical instrumental energy threshold for ACT's is about 200 GeV; however, in some cases imposing a higher threshold in the analysis can improve the sensitivity of FSR searches, as we discuss below.

\subsection{Previous Observations and Upper Bounds}

Previous ACT observations of dwarf galaxies are listed in Table~\ref{observations}. No significant signals were observed, resulting in the listed upper bounds on high energy gamma ray flux from dark matter annihilation in these galaxies. The VERITAS upper limits for the three dwarfs were at the level of 1\% of the Crab nebula~\cite{willman}, which is the value listed in the table. As mentioned earlier, we will not use the Sagittarius bound in our analysis; however, it is worth mentioning that earlier studies have found dark matter interpretations of PAMELA and Fermi anomalies to be incompatible with this bound for optimistic halo profiles of Sagittarius~\cite{lightparticleconstraint}. Fermi LAT telescope has also observed several dwarf galaxies, placing flux upper limits of $\mathcal{O}$$(10^{-10})$~cm$^{-2}$s$^{-1}$ above 1 GeV and $\mathcal{O}$$(10^{-9})$~cm$^{-2}$s$^{-1}$ above 100 MeV; individual bounds are listed in \cite{fermidwarfs}.

\begin{table}[h!]
\begin{center}
\begin{tabular}{|c|c|c|r|} \hline
Dwarf & observed by & total exposure & upper bound on photon flux (cm$^{-2}$s$^{-1}$)\\
 &  &time (hrs) & \\
\hline
Sagittarius & HESS & 11 & $3.6\times10^{-12}$ above 250 GeV \cite{sagittarius}\\
Draco &VERITAS &20 & $2.4\times 10^{-12}$ above 200 GeV \cite{willman}\\
Ursa Minor &VERITAS &20  &$2.4\times 10^{-12}$ above 200 GeV \cite{willman}\\
Willman 1 & VERITAS &15 &$2.4\times 10^{-12}$ above 200 GeV \cite{willman}\\
Willman 1 & MAGIC &15.5 & $10^{-12}$ above 100 GeV \cite{magicwillman}\\
Draco & MAGIC & 7.8 &$1.1\times 10^{-11}$ above 140 GeV \cite{magicdraco}\\
Draco & STACEE & 10.2 & $1.2 \times 10^{-7} (E/{\rm GeV})^{-2.23}$ \cite{staceedraco}\\
Draco & Whipple & 14.3 & $E^2(dF/dE)$~at~400 GeV $< 5.10\times10^{-12}$ erg\cite{whipple}\\
Ursa Minor & Whipple & 17.2 &$E^2(dF/dE)$~at~400 GeV $< 7.30\times10^{-12}$ erg\cite{whipple}\\
\hline
\end{tabular}
\caption{Previous observations of dwarfs by ACTs and corresponding bounds on photon flux} 
\label{observations}
\end{center}
\end{table}

\begin{figure}[h!]
\centering
\includegraphics[width=4.5in,height=2.2in]{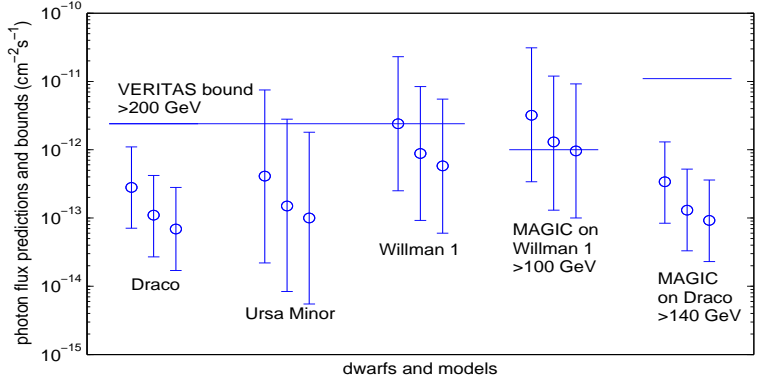}
\caption{Comparison of experimental bounds with predictions from theory. The horizontal lines represent the bounds from Table~\ref{observations}. The three vertical bars for each search are the corresponding predictions of models A (left bar), B (center), and C (right), using the astrophysical factors from Table~\ref{dwarftable}.}
\label{bounds}
\end{figure}

Figure~\ref{bounds} compares these bounds with the predictions of models A, B, C. The signal fluxes are computed using the astrophysical factors in Table~\ref{dwarftable}. We assume the field of view corresponding to $\Delta\theta=0.15^\circ$ in all cases. Experimental bounds from STACEE and Whipple, not shown in the figure, place weaker constraints due to smaller effective areas of the telescopes and lower signal fluxes from Draco and Ursa Minor. Figure~\ref{bounds} shows that the predictions are consistent with the observed null results within the uncertainties in the astrophysical factors. 

\subsection{Backgrounds}

\begin{figure}[t]
\centering
\includegraphics[width=3.5in,height=2.2in]{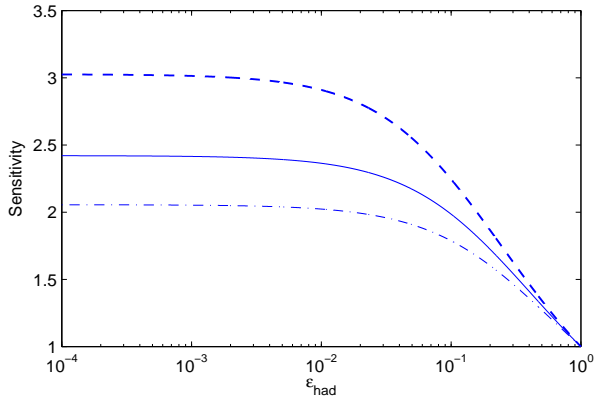}
\caption{Dependence of instrument sensitivity on $\epsilon_{had}$, normalized to 1 at $\epsilon_{had}=1$. The dashed, solid, and dot-dashed lines correspond to energy thresholds of 500, 200, and 100 GeV respectively. }
\label{eps_had}
\end{figure}

Since the dwarf galaxies themselves are not expected to contain significant sources of hard gamma rays of astrophysical origin, the background can be effectively measured by looking at a region of the sky close to the dwarf (called the OFF region). Subtracting the OFF region flux from the flux in the ON region (which contains the dwarf) eliminates the background, up to statistical fluctuations, as long as it is isotropic on the relevant angular scales. If $N$ background photons are expected in the ON region in a given energy bin, the subtraction method effectively reduces the background to $\sqrt{N}$. In order to estimate this residual, which limits the sensitivity of dark matter searches, we use the standard extrapolations of the charged lepton, hadron and gamma-ray cosmic ray spectra~\cite{bergstrom}:
\begin{eqnarray}
\frac{d^2\Phi_{lep}}{dEd\Omega}=1.73\times10^{-8}\left(\frac{100~{\rm GeV}}{E}\right)^{3.3}~{\rm cm}^{-2}~{\rm s}^{-1}~{\rm GeV}^{-1}~{\rm sr}^{-1}\,,\label{hbg}\label{lbg}\\
\frac{d^2\Phi_{had}}{dEd\Omega}=4.13\times10^{-8}\epsilon_{had}\left(\frac{100~{\rm GeV}}{E}\right)^{2.7}~{\rm cm}^{-2}~{\rm s}^{-1}~{\rm GeV}^{-1}~{\rm sr}^{-1}\,,\label{hbg}\\
\frac{d^2\Phi_{\gamma,bg}}{dEd\Omega}\approx3.6\times10^{-10}\left(\frac{100~{\rm GeV}}{E}\right)^{2.7}~{\rm cm}^{-2}~{\rm s}^{-1}~{\rm GeV}^{-1}~{\rm sr}^{-1}\,.\label{gbg}
\end{eqnarray}
Here $\epsilon_{had}$ accounts for rejection of hadronic jets, normalized so that $\epsilon_{had}=1$ for the Whipple telescope. We will conservatively set $\epsilon_{had}=1$ throughout our analysis. Figure \ref{eps_had} shows how instrument sensitivity depends on $\epsilon_{had}$.  Sensitivity improves with hadron rejection, but for sufficiently small values of $\epsilon_{had}$ (below about 0.01) the lepton background starts to dominate, after which the exact value of $\epsilon_{had}$ becomes irrelevant. The figure shows that, with improved hadron rejection, sensitivity can improve by up to a factor of 2 to 3 compared to the sensitivity with $\epsilon_{had}=1$, depending on the threshold energy. 

Note that all backgrounds are assumed to be isotropic. We will ignore the gamma-ray contribution in our analysis since it is significantly smaller than the hadronic and leptonic backgrounds. It is worth keeping in mind that these formulas are simple power law extrapolations from observations at lower energies; we are only relying on these to make estimates,  and actual observations will directly measure and eliminate this background through background subtraction, as mentioned earlier.

In addition to astrophysical background, gamma rays due to dark matter annihilation in the Milky Way should be considered as background for dwarf searches as they contribute approximately 
equally to the flux in the ON and OFF regions. In the leptophilic dark matter models we are interested in, these gamma rays come primarily from inverse Compton scattering (ICS) of starlight and CMB photons off of energetic positrons produced by WIMP annihilation, as well as the FSR process occurring in the Milky Way. We have calculated the ICS distribution using the semi-analytic formalism in~\cite{patrick} and~\cite{positrons}. An isothermal profile for the dark matter density, and the MED propagation model, where the half thickness of the galactic diffusion zone is 4 kpc, have been assumed. Both the ICS and FSR backgrounds depend on the line of sight from the Earth to the respective dwarf and need to be calculated separately for each dwarf. In addition, these backgrounds depend on the dark matter annihilation channel. 

Table~\ref{backgrounds} lists the expected signal and background fluxes, integrated over energy above 200 GeV, for the dwarfs Draco and Segue 1 for each annihilation model. While Draco has the lowest central value for its astrophysical factor among the dwarfs in Table \ref{dwarftable} and is closest in direction to the galactic center, Segue 1 is the exact opposite; hence these provide two extreme cases. The signal fluxes were calculated using the central values of the astrophysical $L$ factors listed in Table~\ref{dwarftable}; in the final column of Table \ref{backgrounds}, we used the astrophysical factor for Segue 1 for an angular resolution of $0.10^{\circ}$, the appropriate parameter for MAGIC-II~\cite{rouven}:
\beq
\log_{10}(L\times \,{\rm GeV}^{-2}{\rm cm}^{5})=20.04\pm1.40.
\eeq{L10}
The WIMP backgrounds are found to be negligible compared to the astrophysical backgrounds for all sources considered here, and will be ignored for the rest of the paper.
Depending on the source, the telescope, and the dark matter model, the signal/background ratios vary between about 1 and 0.005.  For a higher threshold, say 500 GeV, the signal/background ratio will be higher since the background falls off faster than the signal at higher energies. As explained above, the background is measured directly from data, allowing for convincing discovery if sufficient statistics can be accumulated.  

\begin{table}[t]
\begin{center}
\begin{tabular}{|c|c|c|c|}\hline
Dwarf & Draco & Segue 1& Segue 1 \\
&&($\Delta\theta=0.15^{\circ}$)&($\Delta\theta=0.10^{\circ}$)\\
\hline
\underline{signals}&&&\\
Signal Model A & $~2.6\times10^{-13}~$&$~9.8\times10^{-12}~$&$~7.3\times10^{-12}~$\\
Signal Model B & $9.9\times10^{-14}$&$3.7\times10^{-12}$&$2.7\times10^{-12}$\\
Signal Model C & $6.5\times10^{-14}$&$2.4\times10^{-12}$&$1.8\times10^{-12}$\\
\hline
\underline{backgrounds}&&&\\
non-WIMP bg & $2.0\times10^{-11}$&$2.0\times10^{-11}$&$8.8\times10^{-12}$\\
ICS Model A & $3.3\times10^{-16}$&$1.5\times10^{-16}$&$6.8\times10^{-17}$\\
ICS Model B & $4.8\times10^{-16}$&$2.2\times10^{-16}$&$9.9\times10^{-17}$\\
ICS Model C & $2.5\times10^{-16}$&$1.2\times10^{-16}$&$5.2\times10^{-17}$\\
FSR Model A& $2.3\times10^{-15}$ &$1.3\times10^{-15}$&$5.9\times10^{-16}$\\
FSR Model B & $8.5\times10^{-16}$&$4.9\times10^{-16}$&$2.2\times10^{-16}$\\
FSR Model C & $5.6\times10^{-16}$&$3.2\times10^{-16}$&$1.4\times10^{-16}$\\
\hline
\end{tabular}
\caption{Signal and background fluxes (in cm$^{-2}$s$^{-1}$) above 200 GeV for Draco and Segue 1. Dark matter-induced ICS and Milky Way FSR backgrounds are subdominant to astrophysical backgrounds and can be ignored.}
\label{backgrounds}
\end{center}
\end{table}

\section{Detection Prospects}
\label{prospects}

In this section, we will estimate detection prospects of current and near-future ACTs for FSR photons within the three dark matter models at hand. We will assume that the background subtraction method is used, and ignore systematic errors. For positive detection, we impose the requirements that the significance of the excess in the ON region relative to the OFF region exceed 3$\sigma$ (or 5$\sigma$), and the number of photons in the ON region exceed 25:
\begin{eqnarray}
{\mathrm{Significance}}\,=\,\frac{\Phi_{\gamma}A_{\rm eff}t}{(A_{\rm eff}t)^{1/2}(d\Phi_{\rm bg}/d\Omega\times\Delta\Omega)^{1/2}}\geq 3 \mathrm{~(or ~5)},\label{condition1}\\ \nonumber \\
{\mathrm{Number ~of ~signal~events}}\,=\,\Phi_{\gamma}A_{\rm eff}\,t \geq 25,
\label{condition2}
\end{eqnarray}
where $A_{\rm eff}$ is the effective area of the instrument and $t $ is the observation time. For these estimates the angular sizes of the ON and OFF regions were assumed to be equal, and set to the minimum value consistent with the angular resolution.\footnote{Best sensitivity is achieved by minimizing the size of the ON region to improve $S/\sqrt{B}$, as long as the ON region is large enough to capture most of the FSR photons from the dwarf. Increasing the size of the OFF region is not desirable, since angular variations in the background may become an issue, and in any case the residual background is limited by the ON region size.} It is found that for most ACT observations of dwarf galaxies, measurement is limited by backgrounds rather than statistics.

\begin{figure}[h!]
\centering
\includegraphics[width=3.5in,height=2.2in]{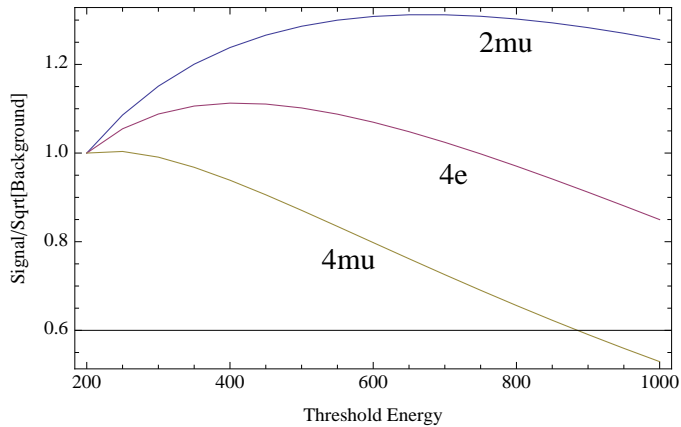}
\caption{$\Phi_{signal}/\sqrt{\Phi_{bg}}$ as a function of threshold energy above which the flux is integrated for the three annihilation channels of interest, normalized to 1 at 200 GeV.}
\label{thresholdoptimize}
\end{figure}

To improve sensitivity, it is useful to choose an energy threshold that maximizes the ratio $\Phi_{\rm signal}/\sqrt{\Phi_{\rm bg}}$. In Figure~\ref{thresholdoptimize} we plot this ratio as a function of threshold energy for the three annihilation channels. The optimum energy threshold lies between 200 and 700 GeV depending on the model. For our estimates we will use two common values, 200 GeV and 500 GeV, for all three models to allow direct comparison between models.

\begin{figure}[h!]
\centering
\includegraphics[width=5.5in,height=2.2in]{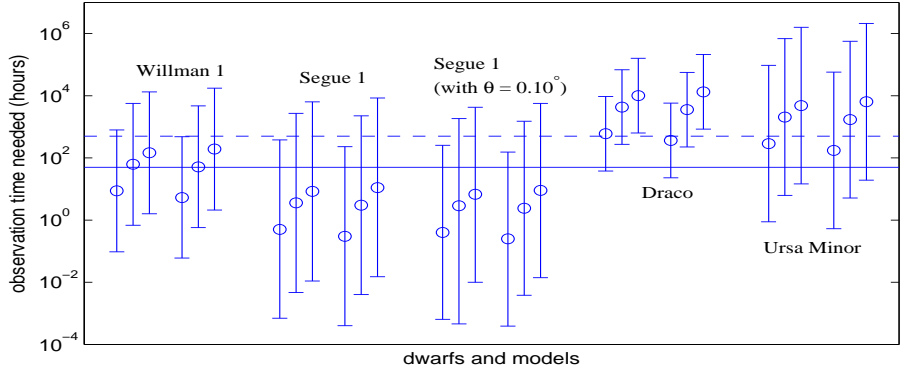}
\caption{Observation times needed for 3$\sigma$ detection with MAGIC parameters. 
The six points for each dwarf correspond to (from left to right) models A, B and C with a 200 GeV energy threshold, and models A, B and C with a 500 GeV threshold. Vertical bars correspond to the uncertainties in the astrophysical factors of the dwarfs from Table~\ref{dwarftable}, with the circles corresponding to the central values from the table. The solid horizontal line denotes 50 hours of observation time. The dashed line, at 500 hours, is equivalent to 50 hours of observation with an order of magnitude increase in the effective area, \textit{i.e.} CTA parameters. }
\label{hoursneeded}
\end{figure}

\begin{figure}[h!]
\centering
\includegraphics[width=4in,height=2.2in]{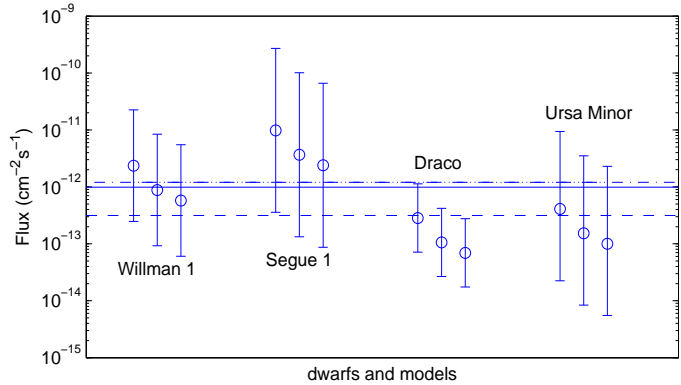}
\caption{Integrated fluxes above 200 GeV. The dot-dashed, solid, and dashed lines correspond to approximate sensitivities of VERITAS, MAGIC, and CTA respectively, for 50 hours of observation time (see Table~\ref{tab:sens}). }
\label{sensitivities}
\end{figure}

\begin{table}[t]
\begin{center}
\begin{tabular}{|c||c|c||c|c|} \hline
ACT & Flux Sensitivity  & Number of events & Flux Sensitivity & Number of events\\
&($>200$ GeV)&($>200$ GeV)&($>500$ GeV)&($>500$ GeV)\\
\hline
VERITAS &$1.2\times10^{-12}$ &150& $5.1\times10^{-13}$&65 \\
MAGIC-II & $9.9\times10^{-13}$&180& $4.3\times10^{-13}$&77\\
CTA &$3.14\times10^{-13}$&565& $1.4\times10^{-13}$&245  \\
\hline
\end{tabular}
\caption{Minimum signal flux (in cm$^{-2}$s$^{-1}$), and the corresponding number of signal events, needed for 3$\sigma$ detection of an excess above the astrophysical background with 50 hrs of observation time. }
\label{tab:sens}
\end{center}
\end{table}

Figure~\ref{hoursneeded} shows the length of observation time needed for a 3$\sigma$ detection of the FSR signal from each dwarf for each annihilation model, for two energy thresholds: 200 and 500 GeV. Likewise, Figure~\ref{sensitivities} shows the approximate minimum integrated flux above 200 GeV that can be detected at the $3\sigma$ level in 50 hours of observation for a few ACTs, and the flux predicted by each model for various dwarfs. These sensitivities are also listed in Table \ref{tab:sens}, along with the corresponding number of events. 

It is clear from the plots that the uncertainties in the astrophysical factors, and therefore observation times and sensitivities required for a positive signal, span several orders of magnitude; as a result, no models are ruled out, and no dwarf is guaranteed to give an observable signal.  Draco has the smallest uncertainty in its astrophysical factor, but is also the least likely to give an observable signal. Segue 1 provides excellent prospects for detection, but the uncertainty on its astrophysical factor is huge. In summary, these estimates only allow us to conclude that, for current and future ACTs, detection of FSR from dark matter annihilation from the above dwarfs is \textit{likely}, but not \textit{guaranteed}. Further astronomical observations of the dwarfs should reduce the uncertainty in the $L$ factors, allowing more precise predictions to be made in the future.

Throughout this section, we conservatively set $\epsilon_{had}=1$, corresponding to the hadron rejection capabilities of the Whipple telescope. Fig.~\ref{eps_had} shows that the sensitivity of the searches can only be enhanced by factors of 2--3 with improved hadronic rejection, before the lepton background becomes dominant. Thus, while certainly desirable, hadronic rejection improvements will not affect the qualitative conclusions of our analysis. 

\section{Model Identification: Some Case Studies}
\label{moddisc}

In this section we investigate the prospects of identifying the dark matter model based on an observed FSR gamma ray signal. As explained above, by ``model" we really mean the final state of WIMP annihilation; to leading approximation used throughout this study, the FSR signal is completely insensitive to details of the microscopic model giving rise to WIMPs and their annihilation. For concreteness, we will focus on the ``models" A, B and C, defined in Section 2. 

Our approach is as follows: We assume that one of the models, with particular parameter values, is realized in nature. (The parameters are $\langle\sigma v\rangle$, $m_\chi$, and, in model C, $m_\phi$. Note that $m_{\phi}$ is not an independent parameter for Model B, where it can be absorbed into $\left<\sigma v\right>$ since there is no additional contribution to FSR from the decay of the final products; see Eq.~\leqn{4state}.) We generate a set of random ``data points" distributed in energy in accord with the theoretical predictions of this model, incorporating the energy resolution of the instrument according to Eq.~\leqn{Eres}. The total number of data points corresponds to the prediction of the model for a particular source and telescope parameters. These data points are then binned, choosing the bin width to be approximately double the energy resolution of the ACT at hand. The WIMP mass will in practice be unknown, but the final bin can always be made large enough that $E_\gamma=m_\chi$ can be assumed to fall in this bin.

For each data set, two background samples, corresponding to the ON and OFF regions, are then generated. The number of background events in each bin is found as the difference between the event counts in the ON and OFF samples in that bin. A fit to the binned data points is performed with all three models\footnote{We use analytic predictions of the models in these fits, ignoring energy resolution of the telescopes. Given the chosen bin widths, energy resolution effects are unlikely to significantly affect the fits in our study. In a more complete analysis, the fit should be performed using Monte Carlo simulations of the signal in each model, which would include energy resolution and other instrumental effects.}, varying the parameters in each model to minimize $\chi^2/d.o.f.$. The three optimal $\chi^2/d.o.f.$ values obtained in this way are compared to each other, and the model with the smallest value is declared the ``best fit" to the data. We perform this procedure for 100 randomly generated data sets, and collect information on the number of data sets for which each of the models (A, B and C) provides the best fit, as well as on the best-fit values of parameters. For most cases, we verified the convergence of model discrimination, as well as the best fit mass and cross section, with a larger number of data sets (about 300).  

We first perform the above analysis assuming that the model A is true, and then repeat it assuming that the ``true" model is B and C. When generating the data set, we fix the model parameters to $m_\chi=3$ TeV, $\langle\sigma v\rangle=1000$ pb, and, where relevant, $m_\phi=1$ GeV. We perform the analysis for two sets of ACT parameter values: $A_{\rm eff}=10^9$ cm$^2$ and 15\% energy resolution, as for MAGIC-II (which we label ``current" below, since these are representative of the reach of current instruments); and $A_{\rm eff}=10^{10}$ cm$^2$ and an energy resolution of 10\%, as for the CTA (which we label ``future"). We use $\Delta\theta=0.10^{\circ}$ for both ON and OFF regions in all cases. 

The results of the analysis are summarized in Tables~\ref{tab:case1}--\ref{tab:case5}. Each table contains the  distribution of ``best fit" models, best fit WIMP mass, cross section, and $\chi^2/d.o.f.$. (Fits to $m_{\phi}$ have very large error bars in almost all cases and therefore are not listed.) The first entry, the distribution of ``best fit" models, contains three numbers, corresponding to the number of data sets for which models A, B, and C respectively gave the best fit. For the other parameters we also list the statistical error bar reflecting the variation of the best-fit values and the $\chi^2/d.o.f.$ of the best fit among the 100 data sets. Discussion of success rates of model identification is based on the assumption that all three models are {\it a priori} equally likely. In practice, information from independent measurements ({\it e.g.} positron spectra) can be used to place priors, enhancing the model discriminating power of the FSR signal.

\vskip0.3cm
\textbf{Case 1: Observation of Segue 1}

\begin{figure}[t]
\centering
\begin{tabular}{cc}
\includegraphics[width=3.1in,height=2in]{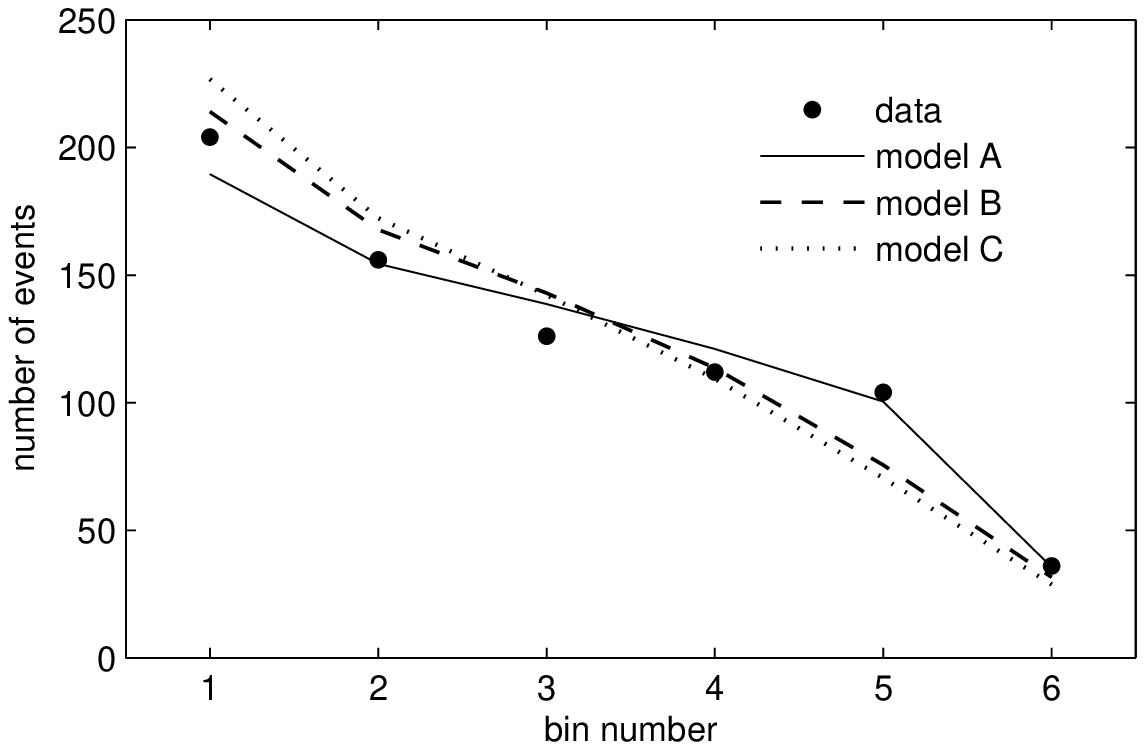}&
\includegraphics[width=3.1in,height=2in]{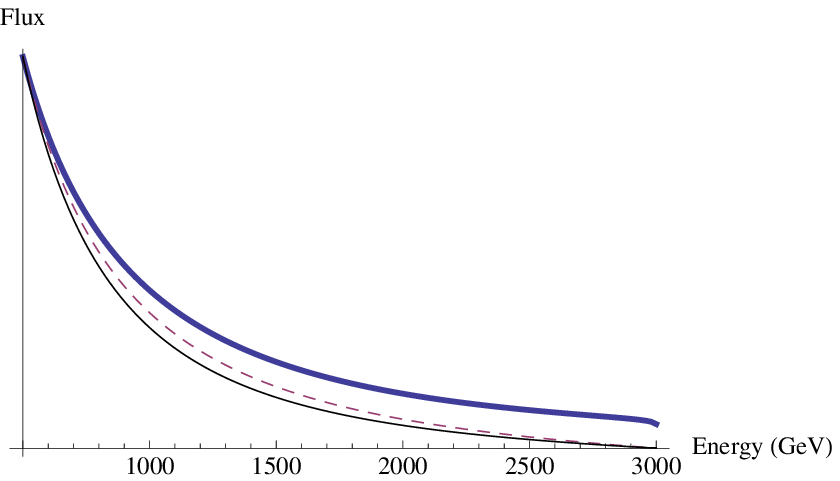}\\
\end{tabular}
\caption{Left: A sample fit. In this particular example, the data is generated according to the model A, which is also correctly identified as the best-fit model. Right: Energy spectra from theory. The thick, dashed, and thin lines correspond to models A, B, and C respectively. The WIMP mass is assumed to be 3 TeV in all models. The flux has been normalized to 1 at 500 GeV for all three curves.}
\label{fig:fitexample}
\end{figure}

As our first, ``benchmark" case study, we consider the observation of the most promising candidate Segue 1, assuming the astrophysical $L$ factor at the current central value from Eq.~\leqn{L10} and an observation period of 50 hours. We consider a data sample with an energy threshold of 500 GeV, distributed in 6 (8) energy bins for current (future) telescope parameters. The results of this analysis are summarized in Table~\ref{tab:case1}, and a sample fit to one of the 100 data sets is shown in the left panel of Fig.~\ref{fig:fitexample}. Note that the cross section errors listed in Table~\ref{tab:case1} and below do {\it not} include the uncertainty of the astrophysical factor $L$. In other words, the errors reflect the relative accuracy of the measured normalization of the signal flux, which is proportional to the product $\sigma \times L$.

\begin{table}[h!]
\begin{center}
\begin{tabular}{|c|c|c|c|c|}\hline
``True" model & model A,B,C & Best fit WIMP  & Best fit  &$\chi^2/d.o.f.$\\
& as best fit & mass (GeV) &cross-section (pb)&\\ \hline
\underline{current}&&&&\\
model A & 100, 0, 0 & $2978\pm112$&$1013\pm34$& $~1.16\pm0.75$~ \\
model B & 16, 44, 40 & $3083\pm412$ & $1054\pm129$ &~ $1.20\pm0.95$~ \\
model C & 17, 13, 70 & $3073\pm449$ & $1199\pm506$& $~1.51\pm1.11$~\\
\hline
\underline{future}&&&&\\
model A & 100, 0, 0 & $3006\pm28$ &$999\pm9$ & $~1.11\pm0.55$~ \\
model B & 0, 68, 32 & $3097\pm144$ & $996\pm38$ &~ $1.37\pm0.89$~ \\
model C & 0, 11, 89 &$3085\pm389$ & $983\pm269$& $~1.42\pm1.05$~\\
\hline
\end{tabular}
\end{center}
\caption{Case 1 fit results.}
\label{tab:case1}
\end{table}

These fits provide several noteworthy results, which also apply to all variations of the analysis considered below (Cases 2--5):

\begin{itemize}

\item The 2$\mu$ channel is very clearly identified. The two 4-body final state channels are not easily distinguishable from each other because their spectra are very similar, see Figure~\ref{fig:fitexample}. The key distinguishing feature of the 2$\mu$ channel is the sharp edge at the end of the spectrum, common to all two-body fermionic channels~\cite{robust}. In contrast, 4-body channels have a softer spectrum, gradually falling off to zero at high energies.

\item Between the 4-body channels, the 4$\mu$ channel is more likely to give a better fit due to the extra degree of freedom in $m_{\phi}$. For this reason, it is often preferred even if the true model is $4e$.

\item Measured values of the WIMP mass and annihilation cross section are in excellent agreement with the ``true" values. The mass is better reconstructed with future telescope parameters because of superior energy resolution, as expected. The accuracy of the mass and cross section determination is particularly impressive (of order 1\%) in the 2$\mu$ channel, again presumably due to the sharp edge feature at $E_\gamma=m_\chi$. 

\item Future telescope parameters provide a clear improvement over current parameters in terms of correct identification of the model. This is mainly due to the increase in effective area, which results in a greater number of events and a better significance.

\end{itemize}

Overall, model identification had a success rate of 71\% for current telescope parameters and 86\% for future parameters in this case. In the cases that follow, we examine the effects of changing some of the assumed parameters.

\vskip0.3cm
\textbf{Case 2: A lower threshold}

To estimate the effect of lowering the energy threshold on model-discriminating power, we repeated the benchmark analysis for a threshold energy of 200 GeV, with data distributed into 9 energy bins (13 for future parameters). All other parameters are the same as in Case 1. 
The results in Table~\ref{tab:case2} show that there is overall improvement in model identification, but the $\chi^2/d.o.f.$ values get slightly worse due to higher background flux at lower energies. Model identification success rates are 79\% and 93\% for current and future telescope parameters respectively. Based on this, there might be some reason to prefer a lower energy threshold, but performing multiple fits with multiple energy threshold cuts on the same signal might be a better approach to identifying the correct model. 

\begin{table}[h!]
\begin{center}
\begin{tabular}{|c|c|c|c|c|}\hline
``True" model & model A,B,C & Best fit WIMP & Best fit  &$\chi^2/d.o.f.$\\
& as best fit & mass (GeV) &cross-section (pb)&\\ \hline
\underline{current}&&&&\\
model A & 99, 1, 0 & $2987\pm107$&$1009\pm26$& $~1.28\pm0.69$~ \\
model B & 12, 59, 29 & $3125\pm372$ & $1024\pm72$ &~ $1.38\pm0.83$~ \\
model C & 5, 16, 79 & $3137\pm436$ & $1062\pm367$& $~1.53\pm0.76$~\\
\hline
\underline{future}&&&&\\
model A & 100, 0, 0 & $3003\pm24$ &$1000\pm7$ & $~1.44\pm0.62$~ \\
model B & 0, 83, 17 & $3074\pm134$ & $997\pm28$ &~ $1.52\pm0.67$~ \\
model C & 0, 3, 97 &$3155\pm347$ & $1018\pm183$& $~1.53\pm0.71$~\\
\hline
\end{tabular}
\end{center}
\caption{Case 2 fit results.}
\label{tab:case2}
\end{table}

\vskip0.3cm
\textbf{Case 3: Improved hadron rejection}

Large atmospheric background is a significant hurdle for dark matter detection from dwarf galaxies using ACTs, as well as for model discrimination. In the benchmark analysis we conservatively set $\epsilon_{had}=1$, corresponding to the hadron rejection capabilities of the Whipple telescope. 
Although we do not have exact values, newer instruments, both current and future, are likely to possess significantly improved hadron rejection capabilities. For this case we consider $\epsilon_{had}=0.01$. According to Figure~\ref{eps_had}, this should improve sensitivity by a factor of ~3, and further reducing $\epsilon_{had}$ does not have much of an effect since the background becomes dominated by the lepton flux. The remaining parameters are the same as in Case 1. The results, listed in Table~\ref{tab:case3}, show some improvement in model identification in all cases . Overall, hadron rejection capabilities of a telescope do not seem to be a limiting factor for model discrimination.

\begin{table}[h!]
\begin{center}
\begin{tabular}{|c|c|c|c|c|}\hline
``True" model & model A,B,C & Best fit WIMP  & Best fit  &$\chi^2/d.o.f.$\\
& as best fit &mass (GeV) &cross-section (pb)&\\ \hline
\underline{current}&&&&\\
model A & 100, 0, 0 & $2999\pm97$&$1005\pm28$& $~1.06\pm0.74$~ \\
model B & 5, 53, 42 & $3082\pm269$ & $1014\pm68$ &~ $0.90\pm0.87$~ \\
model C & 0, 6, 94 & $2951\pm366$ & $575\pm366$& $~1.15\pm0.78$~\\
\hline
\underline{future}&&&&\\
model A & 100, 0, 0 & $3007\pm26$ &$999\pm7$ & $~1.04\pm0.58$~ \\
model B & 0, 74, 26 & $3080\pm83$ & $987\pm20$ &~ $1.06\pm0.59$~ \\
model C & 0, 0, 100 &$3021\pm238$ & $1011\pm184$& $~2.32\pm2.68$~\\
\hline
\end{tabular}
\end{center}
\caption{Case 3 fit results.}
\label{tab:case3}
\end{table}

\vskip0.3cm
\textbf{Case 4: Lighter dark matter}

The benchmark analysis assumed a WIMP mass of  $3$ TeV, which is at the higher end of masses that give good fits to PAMELA, Fermi and HESS data.  While a lower mass results in fewer energy bins and/or comparatively larger background, it also leads to a larger signal due to the $m_\chi^{-3}$ dependence of the FSR signal in Eqs.~\leqn{2state},~\leqn{4state}. To estimate the effect on model discrimination, we repeated the benchmark analysis with $m_{\chi}=1$ TeV, an energy threshold of 200 GeV, and 5 energy bins (7 for future parameters). The results, listed in Table~\ref{tab:case4}, are very similar to the results from case 2, which uses the same energy threshold; the only notable difference is the improved ability to clearly distinguish 2-body channels from 4-body channels with current telescope parameters. Thus, model discriminating capabilities of the ACTs are generally quite robust across the range of WIMP masses preferred by current data within leptophilic models.

\begin{table}[h!]
\begin{center}
\begin{tabular}{|c|c|c|c|c|}\hline
``True" model & model A,B,C & Best fit WIMP  & Best fit  &$\chi^2/d.o.f.$\\
& as best fit & mass (GeV) &cross-section (pb)&\\ \hline
\underline{current}&&&&\\
model A & 100, 0, 0 & $997\pm16$&$1004\pm16$& $~1.02\pm0.77$~ \\
model B & 0, 51, 49 & $1057\pm62$ & $980\pm42$ &~ $1.05\pm0.88$~ \\
model C & 0, 4, 96 & $1064\pm123$ & $997\pm354$& $~1.31\pm1.65$~\\
\hline
\underline{future}&&&&\\
model A & 100, 0, 0 & $1000\pm4$ &$1001\pm4$ & $~1.26\pm0.81$~ \\
model B & 0, 62, 38 & $1021\pm16$ & $985\pm14$ &~ $1.28\pm0.74$~ \\
model C & 0, 0, 100 &$1040\pm62$ & $1031\pm130$& $~1.21\pm0.77$~\\
\hline
\end{tabular}
\end{center}
\caption{Case 4 fit results.}
\label{tab:case4}
\end{table}

\vskip0.3cm

\textbf{Case 5: 3$\sigma$ or 5$\sigma$ detection}

In all studies above, we have used the central value of the astrophysical factor $L$ of Segue 1, which results in rather high fluxes that can be easily detected by MAGIC and CTA, see Fig.~\ref{sensitivities}. 
Due to large uncertainties in the astrophysical factors, in reality fluxes may be much lower. For the lowest possible values of $L$, the ACTs may not be able to even detect the FSR signal from Segue 1; of course, no model discrimination would be possible in this case. It is interesting, however, to examine how much can be learned about the underlying model if the FSR flux is just borderline observable. We performed this analysis, assuming the weakest possible signals that can be detected at 3$\sigma$ or 5$\sigma$ significance. All other parameters are as in case 1. Table~\ref{tab:case5} shows the fit results for current telescope parameters. (Not surprisingly, fit results for future telescope parameters, for the same observation significance levels, are very similar, and therefore not shown.) 

\begin{table}[h!]
\begin{center}
\begin{tabular}{|c|c|c|c|c|}\hline
``True" model & model A,B,C & Best fit WIMP  & Best fit  &$\chi^2/d.o.f.$\\
& as best fit & mass (GeV) &cross-section (pb)&\\ \hline
\underline{3$\sigma$ detection}&&&&\\
model A & 71, 12, 17 & $3092\pm531$&$1235\pm265$& $~1.17\pm0.71$~ \\
model B & 2, 62, 36 & $3041\pm524$ & $1236\pm275$ &~ $1.74\pm1.62$~ \\
model C & 0, 49, 51 & $3051\pm503$ & $1238\pm650$& $~1.26\pm0.77$~\\
\hline
\underline{5$\sigma$ detection}&&&&\\
model A & 83, 9, 8 & $3146\pm433$&$1087\pm182$& $~1.32\pm0.76$~ \\
model B & 2, 64, 34 & $3117\pm471$ & $1148\pm238$ &~ $1.51\pm0.97$~ \\
model C & 3, 17, 80 & $3065\pm489$ & $1005\pm629$& $~1.44\pm0.95$~\\
\hline
\end{tabular}
\end{center}
\caption{Fit results for the weakest possible signal detectable at 3$\sigma$ and 5$\sigma$ levels by the current ACTs.}
\label{tab:case5}
\end{table}

The fit results show that the overall success rate for correctly identifying the model in these scenarios is comparable to the benchmark case 1: 61\% and 75\% for 3$\sigma$ and 5$\sigma$ detection respectively. The success rates of correctly identifying the signal as a 2-body or 4-body final state remain excellent: 90\% and 93\% respectively. Best fit values obtained for the WIMP mass and annihilation cross section remain fairly accurate. We stress again that these results are for the weakest signals that can be detected at the given significance, and any stronger signal will give better fits and model identification.  

\section{Conclusions}

The following is a summary of our main findings:

\begin{itemize}

\item We reviewed the prospects of indirect detection of dark matter via observation of final state radiation (FSR) gamma rays from dwarf galaxies using current and near-future atmospheric Cherenkov telescopes. We found that, assuming dark matter annihilation into leptonic final states, with parameters motivated by the recent PAMELA, Fermi and HESS anomalies, a detection of such a signal is very likely. Unfortunately, lack of precise knowledge of the distribution of dark matter in the dwarfs makes the signal flux predictions highly uncertain, so that null results of these searches would not conclusively rule out these models. (In other words, positive detection is not guaranteed.) 

\item If a signal is observed, fits to the photon spectrum can be used to discriminate between dark matter models. Prospects are especially encouraging for the most promising candidate dwarf Segue 1. With the current central value of its astrophysical factor $L$, our simulations show a 71\% (86\%) success rate of correctly identifying the dark matter model with current (future) telescope parameters. The success rate for determining whether the annihilation is into a 4-body final state or a 2-body final state -- a distinction that can be tremendously helpful for restricting viable models of dark matter -- is close to perfect. 
Distinguishing between 4$e$ and 4$\mu$ channels is more challenging, because their energy spectra are very similar, but future telescopes are expected to provide a significant improvement in this regard.

\item Even if a signal is barely detected at a 5$\sigma$ level, we estimate that there is at least a 75\% chance of correctly identifying the annihilation channel among the three options discussed in this paper. In particular, if annihilation is into two muons, there is about an 83\% chance of correctly identifying it. 

\item Fits to photon energy spectrum can also be used to determine the values of the WIMP mass and annihilation cross section (or, more precisely, $\sigma \times L$). For a signal detected at the 5$\sigma$ level, these parameters can be determined with roughly 15\% accuracy. Note that no prior knowledge of the model is assumed in these fits. 


\item The success rate for proper model identification is fairly robust with respect to changes in energy threshold, WIMP mass, energy resolution, and hadron rejection capabilities of the telescope, and seems to depend mainly on the $S/\sqrt{B}$ ratio. 
\end{itemize}

In summary, if PAMELA, Fermi and HESS anomalies have their origin in leptophilic dark matter annihilation, current and near-future ACTs have an excellent chance of observing the accompanying FSR gamma ray signal from dwarf galaxies. Once a signal is observed, the measured gamma ray spectrum can be used to identify the correct annihilation channel, paving the way to a better understanding of the microscopic nature of dark matter. Our study shows that this approach is very promising. It should be kept in mind, of course, that the phenomenological analysis presented here does not take into account potentially important instrumental effects specific to each telescope, such as instrumental backgrounds, systematic errors on photon energy measurement, energy dependence of the effective area, {\it etc.} We hope that the positive results of our study will encourage experimental collaborations to perform a more careful examination of their instruments' capabilities in this regard, and to continue searches for dark matter annihilation signals from dwarf galaxies.  

\vskip0.5cm
\noindent{\large \bf Note Added} 
\vskip0.3cm
Since the submission of this manuscript, further analysis has driven the astrophysical factor estimates for Segue 1 to lower values \cite{newsegue}; however, we have verified that the general conclusions of our analysis still hold. In particular, the results for Case 5 in Section \ref{moddisc}, which depict worst case scenarios with positive detection, are independent of the value of the astrophysical factor of Segue 1.

\vskip0.5cm
\noindent{\large \bf Acknowledgements} 
\vskip0.3cm

We would like to thank Rouven Essig for useful discussions. This research is supported by the U.S. National Science Foundation through grant PHY-0757868 and CAREER award PHY-0844667.

\end{document}